\documentclass[journal]{IEEEtran}

\usepackage{cite}
\usepackage{graphicx}
\usepackage{tikz}
\usepackage{tikz-3dplot}
\usetikzlibrary{shapes.geometric}
\usetikzlibrary{decorations.pathreplacing}
\usetikzlibrary{positioning}
\usepackage[top=.81in, bottom=.81in, left=.81in, right=.81in]{geometry}  
\usepackage{algpseudocode}
\usepackage[absolute]{textpos}
\usepackage{threeparttable}
\usepackage[strict]{changepage}
\usepackage{bm,upgreek} 
\usepackage{amssymb}
\usepackage{bbm}
\usepackage{hhline}
\usepackage{amsthm}
\usepackage{amsmath}
\usepackage{graphics}
\usepackage{epsfig}
\usepackage{algorithm} 
\usepackage{color}

\ifCLASSINFOpdf

\else

\fi

\usepackage{amsmath}
\usepackage{amsfonts}
\usepackage{amssymb}

\ifCLASSOPTIONcompsoc
\usepackage[caption=false,font=normalsize,labelfont=sf,textfont=sf]{subfig}
\else
\usepackage[caption=false,font=footnotesize]{subfig}
\fi

\usepackage{url}
\usepackage{gensymb}

\begin{document}
	
	%
	
	%
		\title{A Cascaded Multi-IRSs Beamforming Method in mmWave Communication Systems}	
%

	\author{\IEEEauthorblockN{Renjie Liang, Jiancun Fan,~\IEEEmembership{Senior Member,~IEEE}, Yimeng Ge
	}
}


	\maketitle
	
	\begin{abstract}		
In this paper, we study how to jointly design the phase shift of cascaded multi-IRSs and the precoding vector of the BS to improve the coverage in dense urban areas. 
We aim to maximize the signal-to-noise ratio (SNR) of the user equipment (UE) received signal by employing this method.
However, it is a constrained non-convex optimization problem and is NP-hard.
In order to solve this problem, we simplify it by utilizing the characteristic of the mmWave wireless system to decompose the optimization problem into multiple sub-optimization problems. By employing the asymptotic orthogonality of wireless channel in mmWave system to solve the sub-optimization problems, we finally yield a closed-form sub-optimal solution.
The simulation results verify that our solution can effectively improve the coverage of deep dense urban areas.

	\end{abstract}


	%
	\IEEEpeerreviewmaketitle

	\section{Introduction}

Millimeter wave (mmWave) has attracted a lot of attention in future wireless communication system due to huge available bandwidth \cite{revolution}. However, the path loss in mmWave is much larger than that in low frequencies (e.g., sub-6GHz). Especially in dense urban areas, the coverage problem is more serious. Therefore, how to improve the coverage of mmWave system in dense urban areas is an important issue.

\begin{figure}[htbp]
	\centering 
	\includegraphics[scale=0.30]{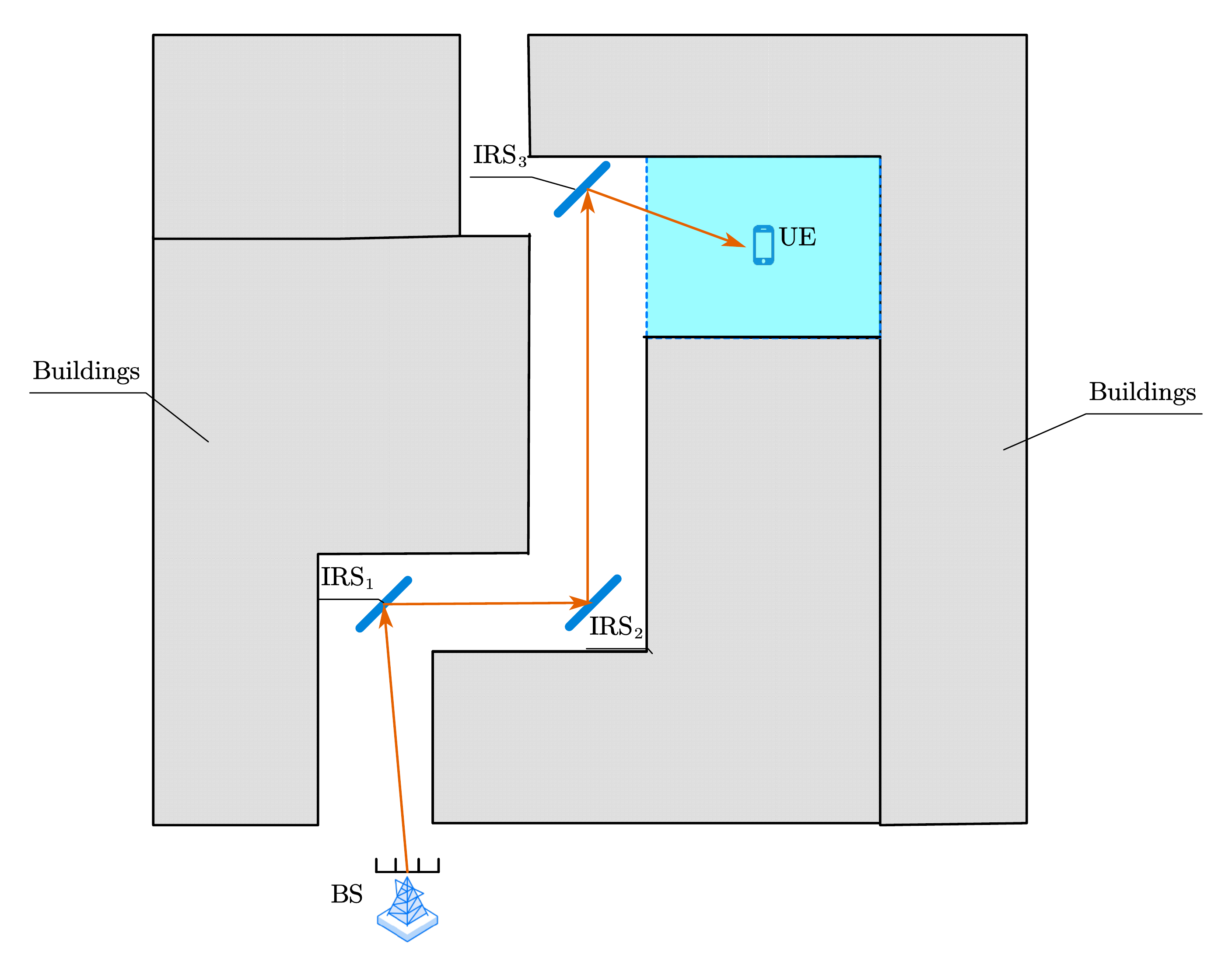}
	\vspace*{-0.1in}
	\caption{The special scenarios of dense urban areas.}
	\label{fig:scenarios}
\end{figure}

In recently years, a new method named intelligent reflecting surface (IRS) is introduced to improve the coverage performance of this scenario, since it can reflect the signal to desired direction through intelligently designing phase shift (PS). 
There have been some related studies employing IRS to improve system performance. As show in Fig.~\ref{}, they mainly focus on jointly optimizing the PS and BS precoding vectors of for three kinds of IRS systems including single IRS, parallel multi-IRSs and cascaded dual-IRSs. For the single IRS, in literature\cite{DBLP}, Q. Wu et al. first propose an semi-definite relaxation (SDR) algorithm to obtain an approximate solution to optimize the PS of a single IRS. However, since the SDR algorithm's computational complexity is too high, they also proposed a distributed algorithm (DA) with a lower complexity. To improve the throughput, Yu et al. \cite{Yu_2019} proposed two other low-complexity algorithms called fixed-point iteration (FPI) and manifold optimization method.
To improve performance for the parallel multi-IRSs system, P. ~Wang et al. \cite{WangIntelligent} proposed an algorithm that can improve the coverage by employing parallel multi-IRSs to assist the BS transmitting data to UEs. 
To improve performance for the cascaded dual-IRSs system, in literature \cite{Propagation Modeling}, Ibrahim. Yildirim et al. introduce a mathematical framework on the error performance for the cascaded double-RISs system.
However, as shown in Fig.~\ref{fig:scenarios}, in some special scenarios of dense urban areas, it is necessary to use cascaded multi-IRSs (three or more) schemes to improve coverage. Moreover, how to optimize the PS of cascaded multi-IRSs (three or more) to improve coverage performance is an important issue that needs to be solved.

As shown in Fig.~\ref{fig:scenarios}, an easy-to-think method is to convert the cascaded multi-IRSs problem into multiple single IRS problems, and employ the optimization algorithm of the single IRS system to solve it. Firstly, we treat the second IRS and subsequent IRS as a single UE. Then, we use a single IRS optimization algorithm to solve the PS of the first IRS and make the PS value fixed. Finally, we adopt a similar method to solve the all PS of the remaining IRS and precoding matrix of the BS. Based on the above results, we can get a sub-optimal solution of received power of the UE.
However, this method is to solve the PS of each IRS individually rather than jointly. Therefore, the solution is not the global optimal, but the local optimal, and its performance is not .

In this paper, different from all the above algorithms, we employ a different method.
we aim to study how to improve the coverage of the dense urban areas by employing a cascade multi-IRSs in a mmWave communication system.  
Our target is to maximize the SNR of the signal received by UE in the dense urban area by jointly optimizing the precoding vector of the BS and the PS of cascade multi-IRSs.
However, this problem is difficult to solve, since it is a constrained non-convex optimization problem and NP hard. 

In order to solve this problem, firstly, we utilize the channel sparsity of mmWave system to convert the problem into a simpler one. 
Secondly, by analyzing the characteristics of the mmWave channel and system model, we clarify that the PS of all IRSs are independent of each other. 
Simultaneously, according to literature \cite{DBLP}, we can know that for the precoding vector of the BS, the maximum ratio transmission (MRT) method is usually employed to obtain a sub-optimal solution. 
Therefore, we can find that the precoding vector at the BS is independent of PS matrix of all the cascaded multi-IRSs. Based on above analysis, we can decompose our problem into multiple simpler sub-problems. Based on the solutions of above sub-problems and considering the asymptotic orthogonality of the array response vector of the BS in mmWave system, we finally obtained a closed form sub-optimal solution. 

Compared with the coverage of the Distributed Algorithm which is introduced in literature \cite{DBLP}, our proposed method has better performance.
The main contributions of this paper are as follows:
	\begin{enumerate}
		\item As shown in Fig.~\ref {fig:RadioEnv}, different from the previous work, we design a cascaded multi-IRS assisted (three or more) scheme to improve the UE received SNR and guarantee the coverage of mmWave wireless systems in dense urban area. 
		\item  Based on the designed scheme, we formulate a constrained non-convex optimization problem, and then employ the channel sparsity of the mmWave system and the progressive orthogonality of the BS's multi-antenna to decompose our problem into multiple simpler sub-problems. Finally, we yield a sub-optimal solution of the optimization problem by solving the sub-problems.
		\item The simulation results show that comparing with DA \cite{} method, our proposed method has better coverage performance .
		
	\end{enumerate}	


The rest of this paper is organized as follows. In Section II, we introduce the radio environment, put forward channel model and derive out the system model. Section III outlines the problem formulation. 
In Section IV, we solve the constrained optimization problem and come up with a sub-optimal solution. Based on this solution, we draw two interesting conclusions. 
In Section V, we simulate the proposed algorithm compared with other methods to prove the advantage of our algorithm. We conclude the paper in Section VI.
\section{System Model and Channel Model}
\label{sec:sysmod}

In this section, firstly, we introduce the radio environment of dense urban area and the channel model, and then we derive the system model.
	\subsection{Radio Environment}
	\label{sec:radio_environment}
	
As shown in Fig.~\ref{fig:IRSRadioEnvPara0903}, we assume that there is a dense urban area with a tortuous path and a time division duplex (TDD) mmWave wireless system is deployed. In this system, a single BS with $N$ antennas is placed outside of the dense urban area. There are $K$ cascade IRSs placed at the corners of outside, every single IRS contain $M$ reflecting elements (REs). The $k$-th IRS is named as $\text{IRS}_k$, where $k \in 1,2,3,\dots,K$, and its PS matrix is $\boldsymbol{\Theta}_k=\text{diag}([\beta e^{j\theta_{k ,1}},\beta  e^{j\theta_{k,2}},\dots ,\beta  e^{j\theta_{k,M}}])$, where $\beta=1$ and $\theta_{k,i} \in(0,2\pi)$, $i=1,2,\dots,M$. There is a single antenna UE stay in blue area of the dense urban area. The distance between the UE and the $\text{IRS}_{K}$ is $d_{r,K}$. The wireless links between the BS and the $\text{IRS}_1$, between the UE and the $\text{IRS}_K$, between every two adjacent IRSs $\text{IRS}_k$ and $\text{IRS}_{k+1}$ are visible. All the rest wireless links are ignored due to severe blocked. Our target is to improve the coverage of this scenario in the dense urban area.


\begin{figure}[htbp]
	\centering 
	\includegraphics[scale=0.3]{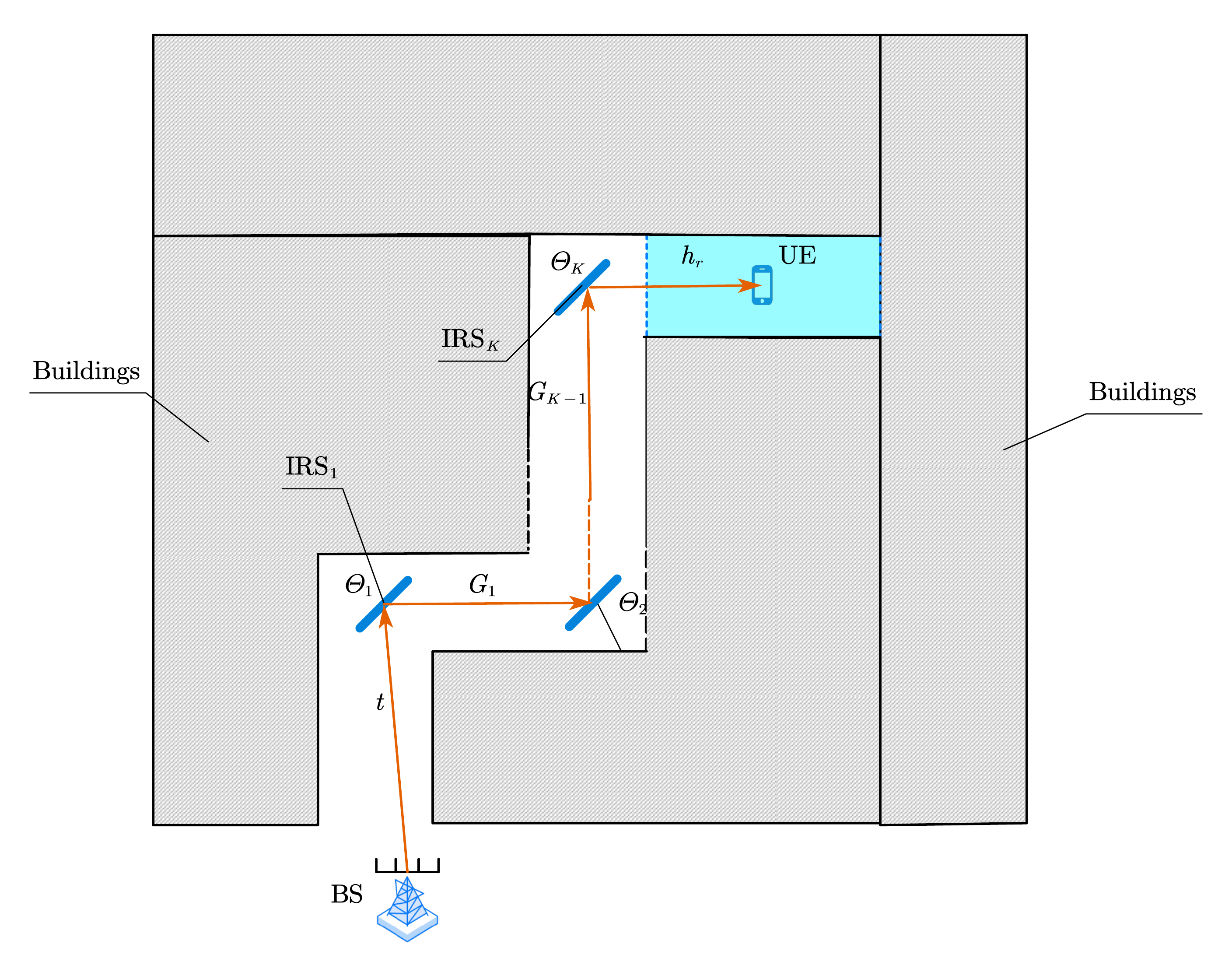}
	\vspace*{-0.1in}
	\caption{The wireless channels in dense urban area with K IRSs.}
	\label{fig:RadioEnv}
\end{figure}

\subsection{Channel Model}
\label{subsec:CM}

As shown in Fig.~\ref{fig:RadioEnv}, let  $\boldsymbol{t}\in \mathbb{C}^{M\times N}$,  $\boldsymbol{G}_{k}\in \mathbb{C}^{M\times M}$, and $\boldsymbol{h}_{r}^H\in \mathbb{C}^{1\times M}$ denote the wireless channel from BS to the first IRS ($\text{IRS}_{1}$), from the $k$-th IRS ($\text{IRS}_{k}$) to $k+1$-th IRS ($\text{IRS}_{k+1}$), and from the $k$-th IRS ($\text{IRS}_{K}$) to UE, respectively, where $k=1,2,\dots,K-1$.

For the mmWave wireless system, many researchers \cite{precoder,precoding,processing,sparse} employ the extended
Saleh-Valenzuela (SV) channel model. In this paper, we also employ this model to derive all above wireless channels.	

According to the extended SV model, the wireless channel $\boldsymbol{G}_k$ between the $\text{IRS}_{k}$ and $\text{IRS}_{k+1}$ can be modeled as
\begin{equation}
\label{eq:GK}
\boldsymbol{G}_k=\sqrt{\frac{M^2}{L}}\sum^L_{l=1}g_{l,k}\Lambda_{t,k}\Lambda_{r,k}\boldsymbol{\alpha}_{r,l,k}(\theta_{l,k})\boldsymbol{\alpha}_{t,l,k}^H(\phi_{l,k}),
\end{equation}
where $g_{l,k}$ denotes the complex path gain of the $l$-th path,  $\boldsymbol{\alpha}_{r,l,k}(\theta_{l,k})$ denotes the normalized receive array response vector associated with the AoA $\theta_{l,k}$, $\boldsymbol{\alpha}_{r,l,k}(\theta_{l,k})\in \mathbb{C}^{M\times 1}$, and the $\boldsymbol{\alpha}_{t,l,k}(\phi_{l,k})$ denotes the normalized transmit array response vector associated with the AoD $\phi_{l,k}$, $\boldsymbol{\alpha}_{t,l,k}(\phi_{l,k})\in \mathbb{C}^{1\times M}$, $\Lambda_t$ and $\Lambda_r$ denote the antenna gain of the transmitter and receiver respectively. 

Compared to sub-6 GHz wireless channels, the number of channel paths $L$ is normally a small number in mmWave wireless channels. For simplicity, we can assume $L=1$, thus (\ref{eq:GK}) can be abbreviated as
\begin{equation}
\label{eq:GK2}
\boldsymbol{G}_k=\sqrt{M^2} g_{k}\Lambda_{t,k}\Lambda_{r,k}\boldsymbol{\alpha}_{r,k}(\theta_{k})\boldsymbol{\alpha}_{t,k}^H(\phi_{k}),
\end{equation}

Let $\mu_{k}=\sqrt{M^2} g_{k}\Lambda_t\Lambda_r$, $\boldsymbol{\alpha_{k}}=\boldsymbol{\alpha}_{r,k}(\theta_{k})$, and $\boldsymbol{\beta}_{k}=\boldsymbol{\alpha}_{t,k}(\phi_{k})$, we can further simplify (\ref{eq:GK2}) as
\begin{equation}
\label{eq:GK3}
\boldsymbol{G}_k=\mu_{k}\boldsymbol{\alpha}_{k}\boldsymbol{\beta}_{k}^H,
\end{equation} 
where  $\boldsymbol{\alpha}_{k}\in \mathbb{C}^{M\times 1}$,  $\boldsymbol{\beta}_{k}\in \mathbb{C}^{M\times 1}$, $k=1,2,\dots,K-1$. 

Following the same method, the wireless channel $\boldsymbol{h}_{r}$ can be modeled as
\begin{equation}
\label{eq:hrk1}
\boldsymbol{h}_{r}=\sqrt{\frac{M}{L}}\sum^L_{l=1}g_{l,r}\Lambda_{t,K}\Lambda_{r,K}\boldsymbol{\alpha}_{t,l}^H(\phi_{l,r}),
\end{equation}
where $g_{l}$ is the complex path gain of the $l$-th path, and $\boldsymbol{\alpha}_{t,l}(\phi_{l,r})$ is the normalized transmit array response vector at transmitter associated with the AoD, $\boldsymbol{\alpha}_{t,l}(\phi_{l,r})\in \mathbb{C}^{M\times 1}$, $\Lambda_t$ and $\Lambda_r$ represent the antenna gain of the transmitter and receiver respectively. 

For simplicity, similar to $\boldsymbol{G}_k$, we can also assume that $L=1$. Thus, (\ref{eq:hrk1}) can be abbreviated as
\begin{equation}
\label{eq:hrk2}
\boldsymbol{h}_{r}=\sqrt{M}g_r\Lambda_{t,K}\Lambda_{r,K}\boldsymbol{\alpha}_{t}^H(\phi_{r}).
\end{equation} 

Let $\mu_K=\sqrt{M}g\Lambda_t\Lambda_r$, and $\boldsymbol{\beta}_K=\boldsymbol{\alpha}_{t}(\phi_{r})$, we can further simplify (\ref{eq:hrk2}) as
\begin{equation}
\label{eq:hrk3}
\boldsymbol{h}_{r}=\mu_K\boldsymbol{\beta}_K^H,
\end{equation} 
where $\boldsymbol{\beta}_K\in \mathbb{C}^{M\times 1}$. 

Following the same method, the wireless channel $\boldsymbol{t}$ also can be modeled as
\begin{equation}
\label{eq:t1}
\boldsymbol{t}=\sqrt{\frac{MN}{L}}\sum^L_{l=1}g_{l,t}\Lambda_{t,0}\Lambda_{r,0}\boldsymbol{\alpha}_{r,l}(\theta_{l,t})\boldsymbol{\alpha}_{t,l}^H(\phi_{l,t}),
\end{equation}
where $g_{l,t}$ denotes the complex path gain of the $l$-th path, $\boldsymbol{\alpha}_{r,l}(\theta_{l,t})$ denotes the normalized receive array response vector associated with the AoA $\theta_{l,t}$, $\boldsymbol{\alpha}_{r,l}(\theta_{l,t})\in \mathbb{C}^{M\times 1}$, the $\boldsymbol{\alpha}_{t,l}(\phi_{l,t})$ denotes the normalized transmit array response vector associated with the AoD, $\boldsymbol{\alpha}_{t,l}(\phi_{l,t})\in \mathbb{C}^{N\times 1}$, $\Lambda_{t,0}$ and $\Lambda_{r,0}$ denotes the antenna gain of the transmitter and receiver respectively. 

For simplicity, similar to $\boldsymbol{G}_k$ and $\boldsymbol{h}_r$, we can also assume that $L=1$. Thus, (\ref{eq:t1}) can be abbreviated as
\begin{equation}
\label{eq:t2}
\boldsymbol{t}=\sqrt{MN}g_t\Lambda_{t,0}\Lambda_{r,0}\boldsymbol{\alpha}_{r}(\theta_{t})\boldsymbol{\alpha}_{t}^H(\phi_{t}),
\end{equation} 
where $\boldsymbol{\alpha}_{r}(\theta_{t})\in \mathbb{C}^{M\times 1}$, $\boldsymbol{\alpha}_{t}(\phi_{t})\in \mathbb{C}^{N\times 1}$. 

Let $\mu_0=\sqrt{MN}g_t\Lambda_t\Lambda_r$, $\boldsymbol{\alpha}_0=\boldsymbol{\alpha}_{r}(\theta_{t})$, and $\boldsymbol{\beta}_0=\boldsymbol{\alpha}_{t}(\phi_{t})$, we can further simplify (\ref{eq:t2}) as
\begin{equation}
\label{eq:t3}
\boldsymbol{t}=\mu_0\boldsymbol{\alpha}_0\boldsymbol{\beta}_0^H.
\end{equation} 

The complex gains $g_k$, $g_r$ and $g_t$ which are mentioned in (\ref{eq:GK2}), (\ref{eq:hrk2}) and (\ref{eq:t2}) are generated according to a complex Gaussian distribution. For simplicity, we employ $g$ to represent all of them as
\begin{equation}
\label{eq:cn}
g\sim \mathcal{CN}(0,10^{-\frac{PL_d}{10}}),
\end{equation}
where $PL_{d}$ denotes the path-loss of distance $d$. Here $d$ represents $d_t$, $d_r$ and $d_{IRS}$ as show in Fig.~\ref{fig:IRSRadioEnvPara0903}, and it can be express as 
\begin{equation}
\label{eq: pl0}
PL_d=PL_{d_0}+10n\log(d)\qquad\text{(dB)},
\end{equation}
where $PL_{d}$ denotes the path loss of the distance $d$ m, $n$ denotes the path-loss exponent, and $PL_{d_0}$ denotes the path loss of reference distance $d_0$. 



\subsection{System Model}


As shown in Fig.~\ref{fig:RadioEnv}, the wireless channel $\boldsymbol{t}$, $\boldsymbol{G}_i$ and $\boldsymbol{h}_r$ is LOS, where $i=1,2,\dots,K-1$. All the rest wireless channels are neglected due to serious blocked, we assume the channel state information (CSI) can be estimated through the method in literature \cite{estimation}. Each IRS can adjust the PS of its REs through the central controller \cite{Principles}.
With above settings, the received signal of UE can be written as
\begin{equation}
y=(\boldsymbol{h}_{r}^H\boldsymbol{\Theta}_K\boldsymbol{G}_{K-1}\boldsymbol{\Theta}_{K-1}\dots \boldsymbol{G}_{1}\boldsymbol{\Theta}_{1}\boldsymbol{t}\boldsymbol{w})\boldsymbol{x}+N_0,
\end{equation}	
where $\boldsymbol{x}$ denotes the transmitting signal of BS, which is independent and identically distributed (i.i.d.) random
variable with zero mean and unit variance. $y$ denotes received signal for the UE. $\boldsymbol{w}$ denotes the precoding vector of the BS, $P$ denotes the transmit power of the BS, $\|\boldsymbol{w}\|_2^2\le P$, $N_0$ denotes the additive complex Gaussian noise with zero-mean and variance $\sigma_n^2$ that the UE received.

\section{Problem Formulation} 
\label{sec:PF}

In this section, based on the system model, firstly, we derive the UE received signal power and SNR, then formulate a constrained optimization problem to maximize the SNR of the signal received by UE.

According to system model, the received power of the UE $P_{\text{RX}}$ can be expressed as
\begin{equation}
\label{eq:rxpower}
	P_{\text{RX}}=|\boldsymbol{h}_{r}^H\boldsymbol{\Theta}_K\boldsymbol{G}_{K-1}\boldsymbol{\Theta}_{K-1}\dots \boldsymbol{G}_{1}\boldsymbol{\Theta}_{1}\boldsymbol{t}\boldsymbol{w}|^2,
\end{equation}

Based on (\ref{eq:rxpower}), we can know the SNR $\gamma$ of the signal received by UE can be expressed as
\begin{equation}
\gamma=\frac{|\boldsymbol{h}_{r}^H\boldsymbol{\Theta}_K\boldsymbol{G}_{K-1}\boldsymbol{\Theta}_{K-1}\dots \boldsymbol{G}_{1}\boldsymbol{\Theta}_{1}\boldsymbol{t}\boldsymbol{w}|^2}{N_0}.
\end{equation}

Our target is to maximize the UE received signal power $P_{\text{RX}}$ by jointly optimizing the precoding vector $\boldsymbol{w}$ at BS and the PS matrix $\boldsymbol{\Theta}_k$ of the cascaded multi-IRSs, subject to the power constraint on the precoding vector at BS and uni-modular constraints on the PS matrix. Mathematically, the optimization problem can be formulated as

\begin{equation}
\begin{aligned}
(\text{P1}):\quad\text{max} &\quad |\boldsymbol{h}_{r}^H\boldsymbol{\Theta}_K\boldsymbol{G}_{K-1}\boldsymbol{\Theta}_{K-1}\dots \boldsymbol{G}_{1}\boldsymbol{\Theta}_{1}\boldsymbol{t}\boldsymbol{w}|^2,\\
\text{s.t}&\quad \|\boldsymbol{w}\|_2^2\le P\\
&\quad \boldsymbol{\Theta}_k=\text{diag}([e^{j\theta_{k,1}},e^{j\theta_{k,2}},e^{j\theta_{k,2}},\dots ,e^{j\theta_{k,M}}])
\end{aligned}
\end{equation}
This optimization problem is non-convex and NP hard, so it is difficult to solve. In next section we will employ the characteristic of mmWave wireless channel to simplify and solve it.

\section{Proposed Method} 
\label{sec:MIR}

In this section, firstly, we employ the characteristic of the mmWave system to simplify and solve the constrained optimization problem which is formulated in section \ref{sec:PF}, and then obtain a sub-optimal solution. 

Before solving this problem, we need to define two variables. The first variable $\boldsymbol{u}_k$ is defined as
\begin{equation}
\label{eq:u_k}
\boldsymbol{u_k}=\boldsymbol{\beta}_{k+1}^* \circ \boldsymbol{\alpha}_k^*,
\end{equation}
where $\boldsymbol{\beta}_{k+1}$ denotes the normalized transmit array response vector of the transmitter of the $k+1$-th IRS, $\boldsymbol{\alpha}_k$ denotes the normalized received array response vector of the receiver of $k$-th IRS, the $\circ$ denotes the element-wise product, $\boldsymbol{\alpha}_{k}\in \mathbb{C}^{M\times 1}$, $\boldsymbol{\beta}_{k}\in \mathbb{C}^{M\times 1}$.
The second variable $\boldsymbol{\theta}_{k}$ is defined as
\begin{equation}
\label{eq:thetak1}
\boldsymbol{\theta}_{k}=[e^{j \theta_{k, 1}},\ldots,e^{j \theta_{k, M}}],
\end{equation}
where the $\boldsymbol{\theta}_{k}$ denotes the PS vector of $\text{IRS}_k$ and $\boldsymbol{\theta}_{k}\in \mathbb{C}^{1\times M}$. Recalling that we have defined a similar variable $\boldsymbol{\Theta}_{k}$ which is the PS matrix of the $k$-th IRS and $\boldsymbol{\Theta}_{k}\in \mathbb{C}^{M\times M}$, we can find the relationship between them can be expressed as
\begin{equation}
\boldsymbol{\Theta}_{k}=\text{diag}(\boldsymbol{\theta}_{k}).
\end{equation}

Substituting (\ref{eq:GK3}), (\ref{eq:u_k}) and (\ref{eq:thetak1}) in problem (P1). After a series of simplifications, we can get the problem (P2):

\begin{equation}
\begin{aligned}
(\text{P2}):\text{max} &\quad |\prod _{i=0}^{K}\mu_i  \boldsymbol{\theta}_K \boldsymbol{u}_K \boldsymbol{\theta}_{K-1} \boldsymbol{u}_{K-1}\dots \boldsymbol{\theta}_1 \boldsymbol{u}_1 \boldsymbol{\beta}_0^T\boldsymbol{\omega}|^2\\
\text{s.t}&\quad \|\boldsymbol{\omega}\|_2^2\le p\\
&\quad \boldsymbol{\theta}_{k}=[e^{j \theta_{k, 1}},\ldots,e^{j \theta_{k, M}}], 
\end{aligned}
\end{equation}

 By investigating the structure of problem (P2), we can find the product of the PS vector $\boldsymbol{\theta}_{k}$ of the $k$-th IRS and $\boldsymbol{u_k}$ is a complex scalar. Therefore, we can reformulate the problem (P2) as
 \begin{equation}
 \begin{aligned}
 (\text{P3}):\quad\text{max} &\quad |\prod_{i=0}^{K}\mu_i \prod_{k=1}^{K}(\boldsymbol{\theta}_k \boldsymbol{u}_k) \boldsymbol{\beta}_0^T\boldsymbol{\omega}|^2\\
 \text{s.t}&\quad \|\boldsymbol{\omega}\|_2^2\le p\\
 &\quad \boldsymbol{\theta}_{k}=[e^{j \theta_{k, 1}},\ldots,e^{j \theta_{k, M}}], 
 \end{aligned}
 \end{equation}
 
 Based on above analysis, we can know that the PS vector $\boldsymbol{\theta}_{k}$ of the $k$-th IRS is determined by $\boldsymbol{u}_k$. Also from (\ref{eq:u_k}), we can know that $\boldsymbol{u}_k$ is determined by $\boldsymbol{\beta}_{k+1}$ and $\boldsymbol{\alpha}_k$. Therefore, the PS of every IRS is independent with each other. Simultaneously, we know that one optimal solution for the precoding vector of BS $w$ is employing the MRT method, which can be described as 
 \begin{equation}
 \label{eq:mrt}
 \boldsymbol{w}^{\star}=\frac{\sqrt{P}(\boldsymbol{\beta}_0^T)^H}{\sqrt{\|\boldsymbol{\beta}_0^T\|_2^2}}.
 \end{equation}
 
 From above analysis, we can find that the problem (P3) can be decomposed into $K+1$ independent sub-problems. These independent sub-problems include $K$ sub-problems as
 \begin{equation}
 \begin{aligned} 
 (\text{P4}):\quad\max _{\boldsymbol{\theta}_{k}} &\quad \boldsymbol{\theta}_{k} \boldsymbol{u}_{k} \\ 
 \text { s.t. } &\quad \boldsymbol{\theta}_{k}=[e^{j \theta_{k, 1}},\ldots,e^{j \theta_{k, M}}], 
 \end{aligned}
 \end{equation}
 and one sub-problem which is the optimization problem of the precoding vector of BS $w$. However, the last sub-problem has been solved by MRT method in (\ref{eq:mrt}).  
 
It is easy to know that under the condition of $\theta_{k, i}=-\arg (u_{k,i})$, the optimal solution of the problem (P4) is a real number of $\sum_{i=1}^{M} |u_{k,i}|$.


Based on above analysis, we can collect all the optimal solutions of sub-problems together, and then we can simplify the objective function as
\begin{equation}
\label{eq:opt}
\begin{aligned}
&\boldsymbol{h}_{r}^H\boldsymbol{\Theta}_K\boldsymbol{G}_{K-1}\boldsymbol{\Theta}_{K-1}\dots \boldsymbol{G}_{1}\boldsymbol{\Theta}_{1}\boldsymbol{t}\boldsymbol{w}^{\star}\\
&=\frac{\sqrt{P}\prod_{i=0}^{K}\mu_i \prod _{k=1}^{K} (\sum_{i=1}^{M} |\boldsymbol{u}_{k,i}|) \boldsymbol{\beta}_0^T (\boldsymbol{\beta}_0^T)^H}{\sqrt{\|\boldsymbol{\beta}_0^T\|_2^2}}.
\end{aligned}
\end{equation}

If the number of BS antennas is large enough, the normalized transmit array response vector $\boldsymbol{\beta}_0$ can be regarded as asymptotically orthogonal \cite{}. We know that the wavelength of mmWave system is very small, and thus the BS's antennas can be a large number. By employing this characteristic of mmWave wireless system, we have 
\begin{equation}
\label{eq:asymptotic_orthogonality}
 \boldsymbol{\beta}_0^T (\boldsymbol{\beta}_0^T)^H\approx 1,
\end{equation}

Substituting (\ref{eq:asymptotic_orthogonality}) in (\ref{eq:opt}), we can finally get a closed-form sub-optimal solution of the optimization problem (P1) which can be expressed as

\begin{equation}
P_{\text{RX}}^{\star}= \frac{P(\prod_{i=0}^{K}\mu_i \prod_{k=1}^{K} (\sum_{i=1}^{M} |\boldsymbol{u}_{k,i}|))^2}{\|(\boldsymbol{\beta}_0^T)^H\|_2^2},
\end{equation}
where $P_{\text{RX}}^{\star}$ denotes the UE received signal power.
Simultaneously, we can get the SNR $\gamma^{\star}$ of the signal received by UE as
\begin{equation}
\label{eq:gamma}
\gamma^{\star}= \frac{P(\prod_{i=0}^{K}\mu_i \prod_{k=1}^{K} (\sum_{i=1}^{M} |\boldsymbol{u}_{k,i}|))^2}{\|(\boldsymbol{\beta}_0^T)^H\|_2^2N_0^2},
\end{equation}

Substituting  (\ref{eq:GK3}), (\ref{eq:t3}) and (\ref{eq:hrk3}) into (\ref{eq:gamma}), we can get
 
 \begin{equation}
 \label{eq:gamma2}
 \begin{aligned}
 &\gamma^{\star}=\\&\frac{M^{2K}N P(g_r g_t \prod_{i=1}^{K-1}g_i \prod _{i=1}^{K+1}\Lambda_{t,i} \Lambda_{r,i}\prod_{k=1}^{K} (\sum_{i=1}^{M} |\boldsymbol{u}_{k,i}|))^2}{\|(\boldsymbol{\beta}_0^T)^H\|_2^2N_0^2}.
 \end{aligned}
 \end{equation}
 
As we know that the modulus of normalized array response vector is 1. Therefore, $\sum_{i=1}^{M} |\boldsymbol{u}_{k,i}|=1$ and $\|(\boldsymbol{\beta}_0^T)^H\|_2^2=1$ . Based on these two equations, we can simplify (\ref{eq:gamma2}) as
  \begin{equation}
 \label{eq:gamma9}
 \begin{aligned}
 &\gamma^{\star}=\\&\frac{M^{2K}N P(g_r g_t \prod_{i=1}^{K-1}g_i \prod _{i=1}^{K+1}\Lambda_{t,i} \Lambda_{r,i}\prod_{k=1}^{K} )^2}{N_0^2}.
 \end{aligned}
 \end{equation} 
 
It should be noted that (\ref{eq:gamma9}) is complicated and contains random variables, we can give three assumptions to simplify it to yield some conclusions. 

Firstly, we assume the $\Lambda_{t,i}$ and $\Lambda_{r,i}$ are 1, where $i=0,1,2,3,\dots,K$.
Secondly, we know $g_r$, $g_t$ and $g_i$ where $i=1,2,\dots,K-1$ are random variables as described in (\ref{eq:cn}). For simplicity, we assume all of them can be expressed as
\begin{equation}
\label{eq:cn2}
g=10^{-\frac{PL_d}{10}},
\end{equation}
where the $PL_d$ is the same as (\ref{eq: pl0}). Similarly, we can also define  $g_0$ as
\begin{equation}
\label{eq:cn3}
g_0=10^{-\frac{PL_{d_0}}{10}},
\end{equation}
where $d_0$ denotes the reference distance. 
At last, we assume that the distance between any neighbor IRSs $d_{IRS}$ is the same with each other.

Base on above assumption, we can simplify (\ref{eq:gamma9}) as
\begin{equation}
\label{eq:gamma3}
\begin{aligned}
\gamma^{\star}_{K}=\frac{M^{2K}N P g_0^{K+1}}{d_r^n d_t^n d_{IRS}^{n(K-1)}N_0^2},
\end{aligned}
\end{equation}
where $\gamma^{\star}_{K}$ denotes the SNR of the signal received by UE within the coverage of the $K$-th IRS. 

\begin{figure}[htbp]
	\centering 
	\includegraphics[scale=0.40]{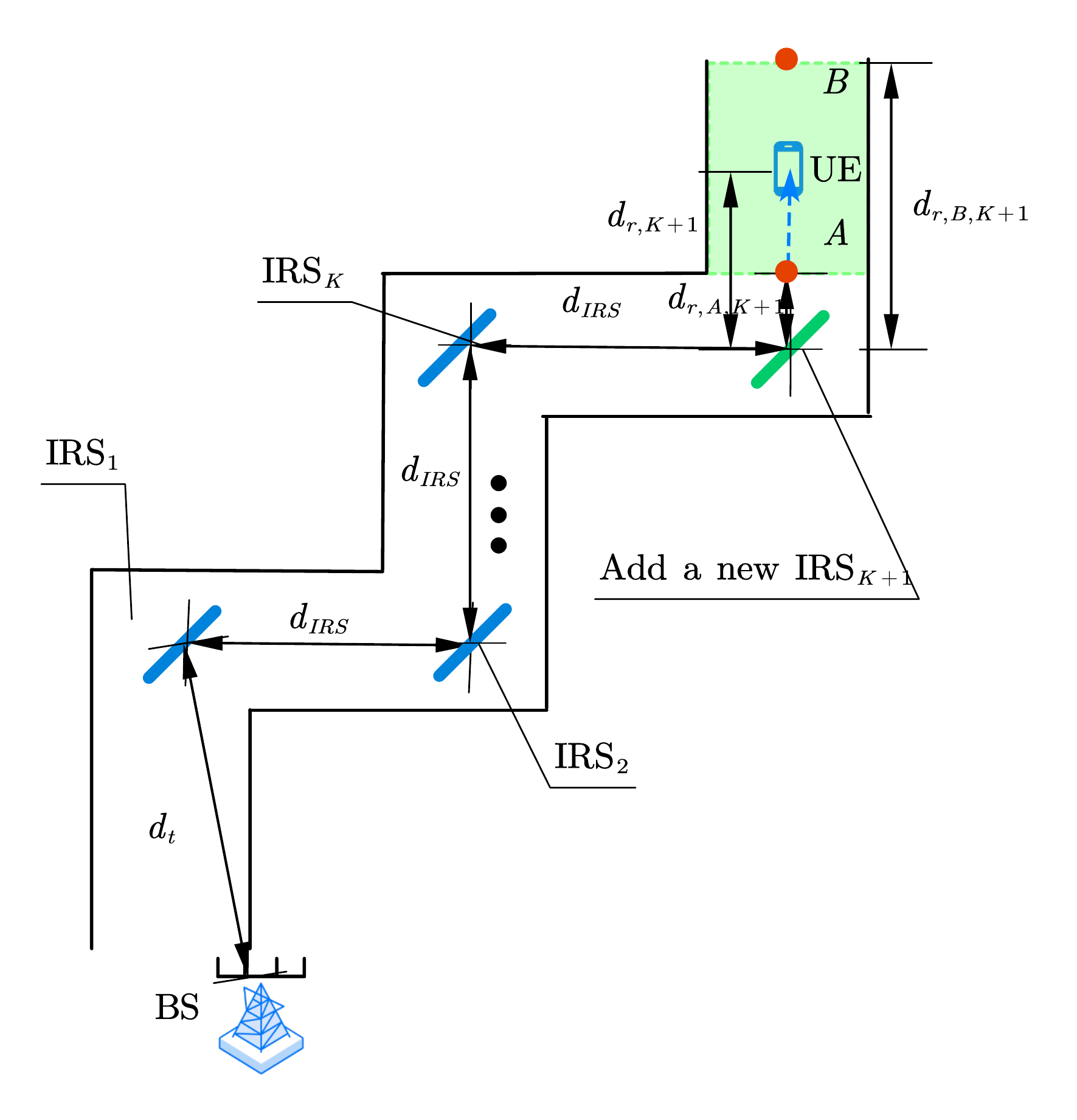}
	\vspace*{-0.1in}
	\caption{The wireless environment of the first scenario}
	\label{fig:sc1}
\end{figure}

As show in Fig.~\ref{fig:IRSRadioEnvPara0903}, a UE is located at the blue area, the distance between UE and the $\text{IRS}_K$ is $d_{r,K}$. Then we deploy a new IRS at the corner $K+1$ as show in Fig.~\ref{fig:sc1}, and move the existing UE to a new location in green area, the distance between the UE and the $\text{IRS}_{K+1}$ is $d_{r,K+1}$. For simplicity, we assume $d_{r,K}=d_{r,K+1}=d_r$.

Based on the same principle, we can know that if add a new IRS, the SNR of the signal received by UE can be express as
\begin{equation}
\label{eq:gamma4}
\begin{aligned}
\gamma^{\star}_{K+1}=\frac{M^{2(K+1)}N P g_0^{K+2}}{d_r^n d_t^n d_{IRS}^{nK}N_0^2},
\end{aligned}
\end{equation}
where $\gamma^{\star}_{K+1}$ denotes the SNR of the signal received by UE at the new location.

Comparing (\ref{eq:gamma3}) with (\ref{eq:gamma4}), we can observe that if we want to let the SNR of the signal received by UE at the new location not less than the original location, we need the number of reflecting elements of each IRS $M$ is larger than $ M_{min}$ which can be express as
\begin{equation}
\label{eq:gamma5}
\begin{aligned}
M_{min}=\sqrt{\frac{d^{n}}{g_0}}.
\end{aligned}
\end{equation} 

From (\ref{eq:gamma5}), we can know if the number of reflecting elements of the IRSs $M$ is smaller than $M_{min}$, it will cause the SNR of the signal received by UE decrease as the deployed IRS increases. In other words, if we want to deploy more IRS without the SNR of the signal received by UE decrease, we need the reflecting elements of IRS to be large enough. 
\section{Simulation Results}\label{sec:results}

In this section, we simulate and compare our proposed method with distributed algorithm (DA) proposed in literature \cite{DBLP}. The simulation results prove the effectiveness of our proposed algorithm.  

\begin{figure}[htbp]
	\centering 
	\includegraphics[scale=0.4]{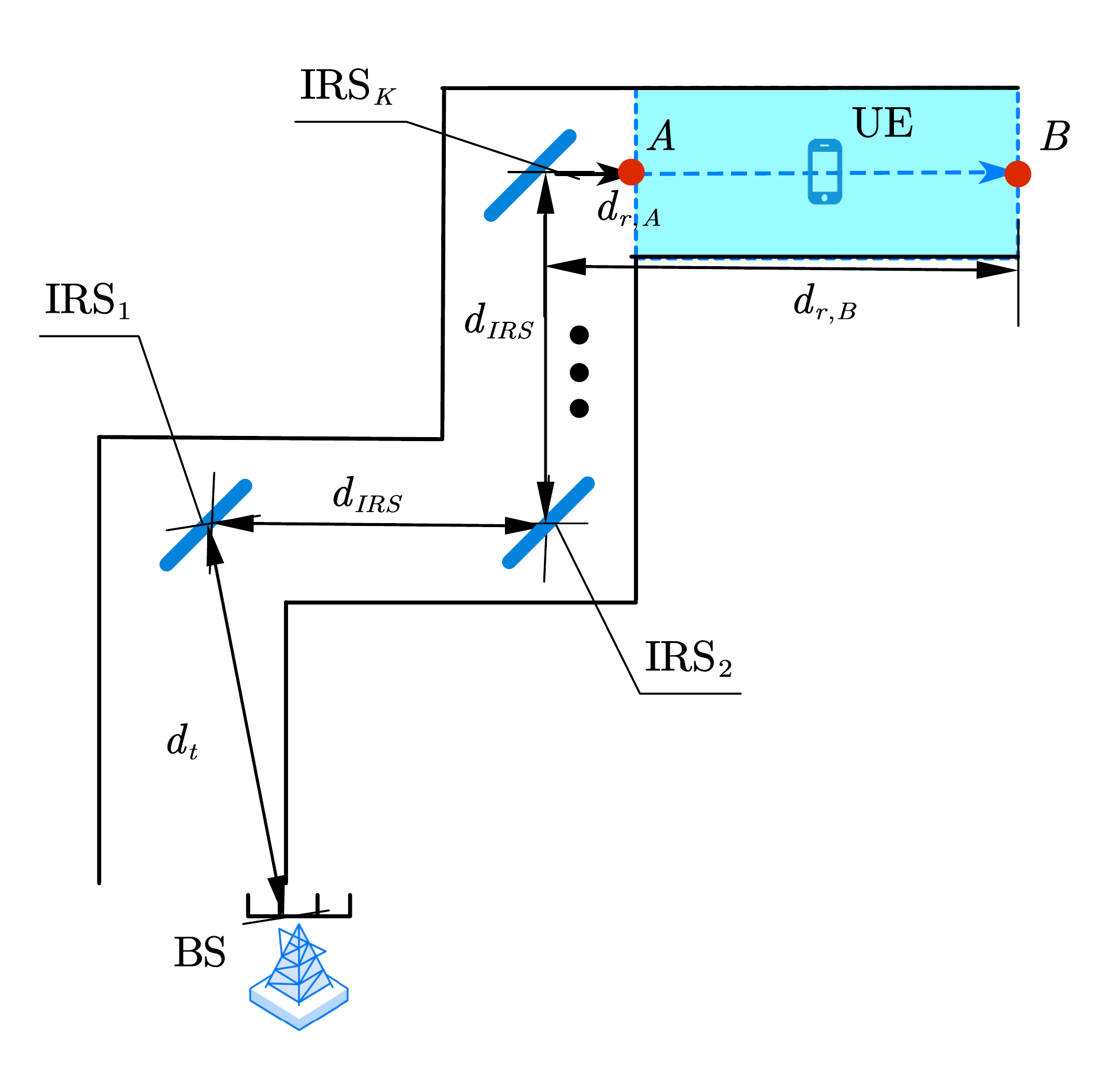}
	\vspace*{-0.1in}
	\caption{The wireless environment in dense urban area with $k$ IRSs.}
	\label{fig:IRSRadioEnvPara0903}
\end{figure}

For the simulation, we consider the frequency of the mmWave system is 28 Ghz, and thus its wavelength $\lambda$ is 0.01 m. we assume the reference distance $d_0$ is 1 m, and its path-loss $PL_{d_{0}}$ is $(\frac{\lambda}{4\pi})^2=-61.4$ dBm for the LOS scenario. The transmit power of BS is 46 dBm, and its antenna number is 128. The additive complex Gaussian noise power of the UE received is -94 dBm.  As show in Fig.~\ref{fig:IRSRadioEnvPara0903}, we assume there is a single antenna UE moving from point A to point B along the dotted line in blue area of the dense urban area. The distance between the UE and the $\text{IRS}_{K}$ is $d_{r,K}$. The distance between the point A and the $\text{IRS}_{K}$ is $d_{r,A}$, and the distance between the ending point B and the $\text{IRS}_{K}$ is $d_{r,B}$. Based on above definitions, we can know $d_{r,K} \in [d_{r,A},d_{r,B}]$. We set $d_{IRS}=20$ m, $\text{IRS}_1$ $d_t=20$ m, $d_{r,A}=1$ m, and $d_{r,B}=100$ m.
Based on above parameters and (\ref{eq:gamma5}), we can yield $M_{min}$ is nearly $5.8*10^4$, which is a very large number.

\begin{figure}[htbp]
	\centering 
	\includegraphics[scale=0.55]{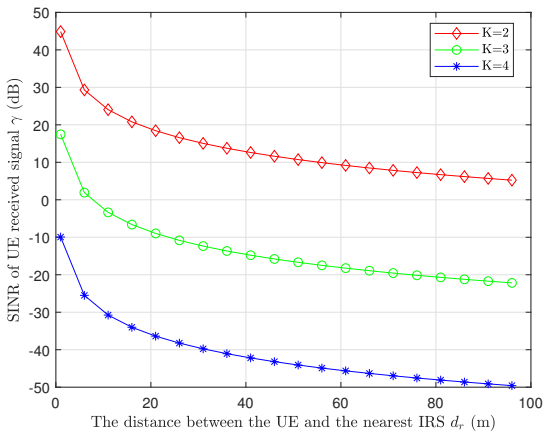}
	\vspace*{-0.1in}
	\caption{The more cascaded multi-IRSs $K$, the worse SNR of the signal received by UE.}
	\label{fig:IRSk}
\end{figure}

As shown in Fig.~\ref{fig:IRSk}, we assume the number of REs of each IRS $M$ is 1000.
We plot the SNR $\gamma$ of the signal received by UE versus the distance $d$ between the UE and $\text{IRS}_K$, under various number of cascaded multi-IRSs $K$.
We can observe that as the number of cascaded multi-IRSs $K$ increases, the received SNR of UE $\gamma$ gradually increases, since $M$ is much smaller than $M_{min}$.

\begin{figure}[htbp]
	\centering 
	\includegraphics[scale=0.55]{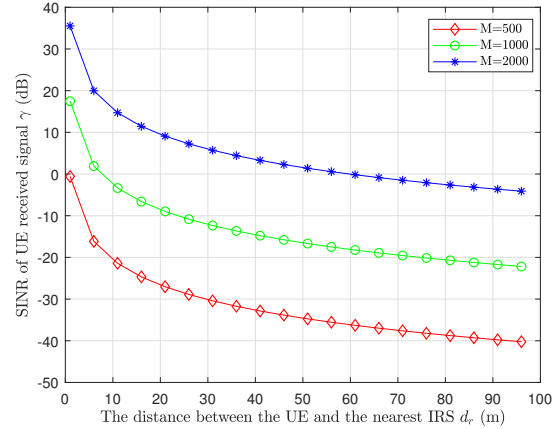}
	\vspace*{-0.1in}
	\caption{The more REs of IRS $M$, the better SNR of the signal received by UE.}
	\label{fig:diffM}
\end{figure}

As shown in Fig.~\ref{fig:diffM},  we assume number of cascaded multi-IRSs $K$ is 3.
we plot the SNR $\gamma$ received by the UE versus the distance $d$ between the UE and $\text{IRS}_K$, under various numbers of the REs $M$ of each IRS.
We can observe that as the number $M$ of the REs of IRS increases, the received SNR $\gamma$ of UE gradually increases.

\begin{figure}[htbp]
	\centering 
	\includegraphics[scale=0.55]{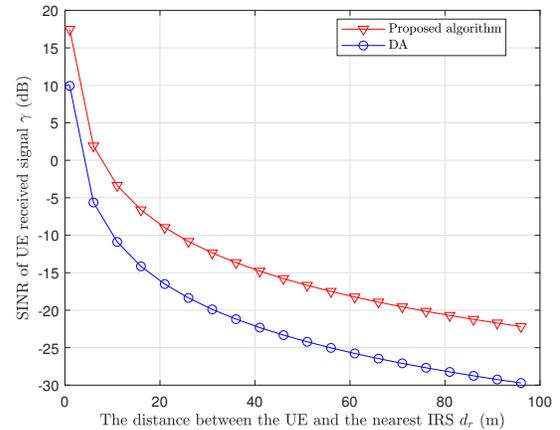}
	\vspace*{-0.1in}
	\caption{Comparing with the NLOS scenario, our proposed method show better coverage performance.}
	\label{fig:IRSnum}
\end{figure}

As shown in Fig.~\ref{fig:IRSnum}, we assume the numbers of cascade IRSs $K$ is 3 and the REs of each IRS $M$ is 1000. We compare our proposed with a extend version  of DA proposed in literature \cite{DBLP}. The DA is a low complexity algorithm to jointly design the PS at IRS and precoding vector at the BS for single IRS assisted point-to-point multiple-input single-output (MISO) wireless system. Therefore, it cannot be benchmarked with our proposed algorithm. Thus, we extend it to the cascaded multi-IRSs system. We plot the SNR $\gamma$ of the received SNR of UE versus distance $d_r$ between the UE and the $\text{IRS}_K$. We can observe that our proposed algorithm has a better coverage performance than DA.

\section{Conclusion}\label{sec:conclusions}

In this paper, we focus on the mmWave system's coverage improvement in dense urban areas by employing cascade multi-IRSs scheme. After formulating a constrained optimization problem to maximize the received SNR of UE, we finally solve it and yield a sub-optimal solution. 
The simulation result shows that our method has better coverage performance than DA. 

\ifCLASSOPTIONcaptionsoff
\newpage
\fi


\bibliographystyle{IEEEtran}

\end{document}